\documentclass[conference]{IEEEtran}

\usepackage{url}
\usepackage[noadjust]{cite}
\usepackage{graphicx}
\usepackage{tabularx}
\usepackage{float}
\usepackage{amsmath}

\usepackage{customcommands}

\begin{document}

\newcommand{\SMA}{\textsc{sma}\xspace}
\newcommand{\DMA}{\textsc{dma}\xspace}
\newcommand{\ADMA}{\textsc{a-dma}\xspace}

\title{Share and Disperse: How to Resist Against Aggregator
  Compromises in Sensor Networks}

\author{
\authorblockN{Thomas Claveirole$^1$, Marcelo Dias de Amorim$^1$, Michel
Abdalla$^2$, and Yannis
Viniotis$^3$\vspace*{2mm}}
\begin{tabular}{c c c}
$^1$ LIP6/CNRS &
$^2$ D{\'{e}}partement d'Informatique &
$^3$ Department of ECE \\
Universit{\'e} Pierre et Marie Curie~-- Paris VI &
{\'{E}}cole Normale Sup{\'{e}}rieure &
North Carolina State University\\
Paris, France &
Paris, France &
Raleigh, NC, USA\\
{\small \tt \{claveiro,amorim\}@rp.lip6.fr} &
{\small \tt michel.abdalla@ens.fr} &
{\small \tt candice@ncsu.edu}\\
\end{tabular}}

\maketitle

\thispagestyle{plain}

\begin{abstract}
A common approach to overcome the limited nature of sensor networks
is to aggregate data at intermediate nodes. A challenging issue in
this context is to guarantee end-to-end security mainly because
sensor networks are extremely vulnerable to node compromises. In
order to secure data aggregation, in this paper we propose three schemes
that rely on multipath routing. The first one guarantees data
confidentiality through \textit{secret sharing}, while the second
and third ones provide data availability through \textit{information
dispersal}. Based on qualitative analysis and implementation, we show
that, by applying these schemes, a sensor network can achieve data
confidentiality, authenticity, and protection against denial of
service attacks even in the presence of multiple compromised nodes.
\end{abstract}

\IEEEpeerreviewmaketitle

\section{Introduction}

Wireless sensor networks (WSN) are computer networks dedicated to
monitoring physical conditions with the help of sensor
nodes~\cite{akyildiz.comnet02}. They support a wide range of
applications including environmental and wild-life monitoring,
building security and home automation, traffic flow measurement,
medical care, and military operations.

In many applications of WSN, data may be sensitive to external
events that are not expected to happen under normal operation of the
network. In particular, data confidentiality and availability are
important characteristics the network should be able to assure.
Guaranteeing such characteristics is a tough task, especially when
the sensor nodes are composed of inexpensive devices with limited hardware
capabilities.\footnote{It is important to note that sensors in WSN
are not necessarily limited in resources, although most problems
become particularly challenging in such a case.} In this case, where
providing tamper resistance is almost impractical, compromising a node is an
easy and attractive option for attackers.

The limited nature of sensor nodes opens up possibilities for
multiple vectors of attack. Provided that radio communication is
expensive in terms of energy consumption, it is very important to
reduce the communication overhead.\footnote{Transmitting 1Kb at a
distance of 100 meters costs as much as executing 3 million
instructions with a general purpose
processor~\cite{pottie00wireless}.} An interesting approach to
achieve such an objective is to perform {\it data aggregation},
where relaying nodes exploit the distributed nature of the network
and perform in-network processing. Guaranteeing security in
aggregation schemes is particularly challenging because node
compromises in such a scenario are doubly problematic, both in terms
data confidentiality (eavesdropping) and availability (denial of
service).  Indeed, by compromising an aggregator node\footnote{That
is, capturing an aggregator node and having  access to its internal state and
cryptographic material.  The attacker may therefore turn an
authorized node into a malicious one.} the attacker would endanger
all of the readings that are part of the aggregate the node is in
charge of.

Several researchers have already studied the problem of securing data
aggregation. Mykletun {\it et al.}~\cite{mykletun05efficient} suggest
using ciphers for which some arithmetical operations over ciphertexts
have some arithmetical signification on the cleartext.  While this
technique allows for some security, a compromised node may still stop
aggregating and forwarding data.  Even worse, tampering and replay
attacks cannot be detected with such a solution. Przydatek {\it et
  al.}~\cite{przydatek03sia} propose a number of techniques to ensure
the integrity of the aggregated data for some aggregation functions.
Although integrity can be satisfactorily assured, the proposed schemes
are difficult to implement and provide neither confidentiality nor
protection against denial of service (DoS) attacks. Hu and
Evans~\cite{hu03secure} propose a scheme that provides authentication
and integrity which is secure even when some nodes are compromised,
however it fails in the case where two consecutive aggregators are
compromised. Furthermore, this scheme neither addresses
confidentiality nor availability.  Wagner~\cite{wagner04resilient}
studies the inherent security of some aggregation functions.  But he
only considers the level of impact a compromised sensor may have on
the final result. His work concerns the security of aggregation
functions, not the aggregation security itself.

In this paper, we do not address data integrity as an explicit issue.
Instead, we focus on confidentiality and availability, which we
believe still lack efficient solutions. To this end, we propose,
analyze, and evaluate three new schemes, namely (a) {\it Secret
  Multipath Aggregation} (\SMA), (b) {\it Dispersed Multipath
  Aggregation} (\DMA), and (c) {\it Authenticated Dispersed Multipath
  Aggregation} (\ADMA). The main idea behind our three approaches is
to exploit  using multiple paths toward the
sink. In fact, a sensor may split a handful of its readings into $n$
separate messages such that $t$ messages are needed to reconstruct the
readings.  By sending messages along disjoint paths, a sensor ensures
that intermediate nodes do not have complete knowledge of the sensed data.
In such a scenario, \SMA guarantees confidentiality by applying the
concept of secret sharing~\cite{shamir79how}. \DMA and its
authenticated version, \ADMA, address availability by dispersing
information over the different paths~\cite{rabin89efficient}.
Although they have been recognized in many research areas ({\it e.g.}, parallel
computing, distributed storage, databases, and ad hoc networking),
surprisingly neither secret sharing nor information dispersal have
been applied to the context of wireless sensor networks nor to the
specific problem of data aggregation.

The  remainder of this paper is organized as follows.
In Section~\ref{sec:prob-form}, we describe the security and network
assumptions considered in the paper. In
Section~\ref{sec:prop-schemes}, we introduce our proposed schemes.
In Section~\ref{sec:sec-ana}, we analyze their security levels.
In Section~\ref{sec:further}, we provide further investigation on the
three schemes and compares them to other approaches. Finally,
in Section~\ref{sec:concl}, we conclude the paper and present some open
issues.

\section{Problem formulation}
\label{sec:prob-form}

In the following, we describe the problems, goals, and assumptions
addressed in this paper.  The section is composed of three parts: (a)
security aspects, (b) network assumptions, and (c) node assumptions.

\subsection{Security goals and threats}
\label{subsec:sec-goals-threats}

The goal of this paper is to provide aggregation schemes that are
resilient to node compromises.  That is, a compromised node alone
should not be able to eavesdrop, tamper, or forbid other nodes from
accessing data.  This paper assumes that resistance against these
attacks in the absence of node compromise is ensured by link-level
mechanisms~\cite{karlof04tinysec, du03pairwise}.

Even a single compromised aggregator node presents a serious threat
to a sensor network's security.  Therefore, some schemes must be
designed to ensure reasonable security in the presence of
compromised aggregator nodes.  Ideally, one would like the network
security to degrade gracefully with the number of compromised nodes.
By security, in this article we mean resistance against the following
attacks: eavesdropping, data tampering, packet injection, and denial
of service.  Other attacks are out of the scope of this paper.

\vspace{1mm} \noindent {\bf Eavesdropping.} Eavesdropping occurs
when an attacker compromises an aggregator node and listens to the
traffic that goes through it without altering its behavior. Since an
aggregator node processes various pieces of data from several nodes
in the network, it does not only leak information about a specific
compromised node, but from a group of nodes.

\vspace{1mm} \noindent {\bf Data tampering and packet injection.} A
compromised node may alter packets that go through it.  It may also
inject false messages.  Since an aggregate message embeds
information from several sensor nodes, it is more interesting for an
attacker to tamper with such messages than simple sensor readings.
An attacker that controls the meaning of the malicious messages it
sends may heavily impact the final result computed by the sink.

An attacker that does not control the meaning of the malicious
messages (for example, if these messages are expected to be
encrypted with a key unknown to the attacker) still can do some
harm.  It may send meaningless garbage values and thus render the
network unusable~-- this is also a form of \DOS attack. Finally, a
particular type of packet injection consists of replay attacks,
where a malicious node eavesdrops some packets in order to re-send
them later.

\vspace{1mm} \noindent {\bf Denial of service.} A compromised node may
stop aggregating and forwarding data.  Doing so, it forbids the data
sink from getting information about several nodes in the network.  If
the node still exchanges routing messages despite its unfair behavior,
that problem may be difficult to solve. The compromised aggregator may
in this way render the network unusable.  Smarter attacks also involve
dropping messages randomly.  It is also difficult to detect when an
attacker sends garbage messages. Finally, it is interesting to note
that such attacks do not necessarily involve a high cost or extended
skills.  For example, a basic \DOS attack may consist of simply
physically breaking the device.

\subsection{Network assumptions}

We assume that each sensor disposes of multiple paths toward the
sink and has link-level encryption capabilities. A node can then
split a flow into several distinct sub-flows and send each one of
them securely toward the sink. Due to encryption, a node cannot
eavesdrop a sub-flow unless it belongs to the path for this flow.

In order to get multiple paths to the data sink, a solution is to
use a multipath routing protocol or disperse several sinks
geographically and  communicate using fast and secure
links.  Ganesan {\it et al.}~\cite{ganesan01highly} study the
establishment of multiple paths in sensor networks and Dulman {\it
et al.} \cite{dulman03tradeoff} explore the relationship between
the amount of traffic and reliability.  Note that schemes described
in this paper require disjoint multipaths to enforce optimal
security. Non-disjoint multipaths may be used, but optimal security
cannot be guaranteed.

We also assume that the underlying routing protocol is secure. In
particular, attention must be paid to spoofing and Sybil
attacks~\cite{karlof03secure}. Roughly speaking, this means that a
node should not be able to impersonate another node or to pretend to
be two distinct nodes. This should not be a problem however if the
link-level encryption keys are distinct amongst the nodes.

\subsection{Node computational/memory assumptions}
\label{sec:prob-form-nodes}

We assume that nodes have very limitedcomputation, memory and
storage capabilities.  This makes many cryptographic algorithms and
protocols impractical, if not impossible to use.  The proposed
schemes were designed to work under such constraints.

We implemented our schemes using the typical Crossbow \MICAz
mote~\cite{micaz}. It uses an Atmel ATmega128L micro-chip (8-bit CPU
at 8~MHz) with 4~Kbytes of RAM and 128~Kbytes of flash memory to
store code and pre-computed program data.  Its energy is provided by
two AA batteries ($\sim$~3V).  It communicates using a 2,4~GHz
IEEE~802.15.4 RF transceiver.

\section{Resisting against aggregator compromises: Proposed schemes}
\label{sec:prop-schemes}

\vspace{1mm} \noindent {\bf Preliminaries.} In this section we present
three schemes to achieve secure aggregation in sensor networks:
\emph{Secret Multipath Aggregation} (\SMA), \emph{Dispersed
Multipath Aggregation} (\DMA), and \emph{Authenticated Dispersed
Multipath Aggregation} (\ADMA).  Each of these schemes has its own
specific characteristics. \SMA offers strong confidentiality at the
cost of some communication overhead. \DMA is optimal with respect to
radio communications but provides a little bit lower level of
confidentiality.  \ADMA adds authentication to \DMA also at the cost
of a slight overhead. All these properties are quantified and
analyzed in sections~\ref{sec:sec-ana} and~\ref{sec:further}.

\begin{figure}
  \centering
  \includegraphics[width=\linewidth]{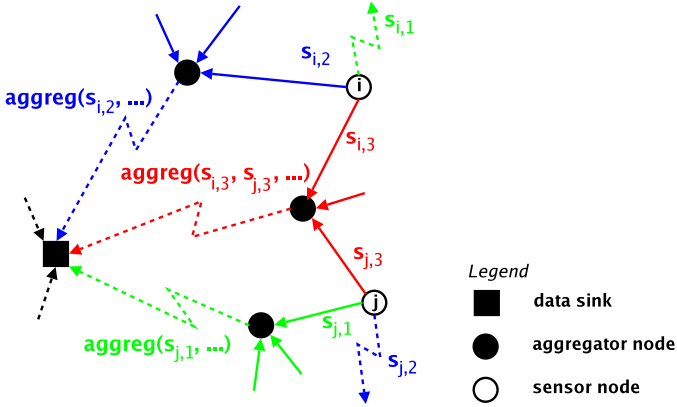}
  \caption{\emph{Proposed schemes basics.} Nodes $i$ and $j$
    respectively split their readings into shares $s_{i, 1}, s_{i, 2},
    s_{i, 3}$ and $s_{j, 1}, s_{j, 2}, s_{j, 3}$.  Shares are then sent
    and aggregated on distinct paths.}
  \label{fig:basics}
\end{figure}

\vspace{1mm} \noindent {\bf Basics.} All the three proposed schemes
use the same basic principle: a sensor node splits its readings into
several \emph{shares} and sends these shares over distinct paths.
Each share makes its way to the data sink. During forwarding, a
share may be processed by aggregator nodes. Once the sink has
gathered enough shares for a given set of readings, it can then
reconstruct this specific set of readings. However, a share alone is
not intelligible to an intermediate node.  Figure~\ref{fig:basics}
depicts this.

\vspace{1mm} \noindent {\bf Tolerance to losses.} The way shares are
constructed depends on the scheme ({\it e.g.}, \SMA's encode only
one reading per share while both \DMA and \ADMA encode multiple
readings per share). The number of shares transmitted and the number
of shares required for reconstructions are not necessarily equal,
which means that the system tolerates some losses during forwarding.


\vspace{1mm} \noindent {\bf Security implications.} The
abovementioned properties yield two interesting security
implications.  First, an attacker must compromise many nodes to be
able to reconstruct readings. This ensures confidentiality. Second,
malicious nodes that stop forwarding shares have limited impact on
the system, since another subset of shares may be used to
reconstruct readings. This ensures protection against \DOS attacks.


\vspace{1mm} \noindent {\bf Homomorphism.} A key point of these
schemes is their homomorphic properties, {\it i.e.}, the ability for
aggregator nodes to perform computations on shares despite their
unknown meaning.  Say nodes $i$ and $j$ sense $r_i$ and $r_j$.  An
aggregator node may add up two shares from $i$ and $j$, which gives
a corresponding share $r_i + r_j$.  This holds for several
aggregation functions on the shared secret, such as sum, mean,
variance, and count~\cite{mykletun05efficient, wagner04resilient}.

\subsection{Scheme 1: Secret Multipath Aggregation (\SMA)}

\SMA applies secret sharing to create shares, which is a common
approach when dealing with security under the contingency of node
compromise.

\vspace{1mm} \noindent {\bf Share creation.} Assume a node $i$ may
use $p$ distinct paths to reach the sink, $t - 1$ ($1 \leq t \leq p$)
of which may be compromised ({\it i.e.}, a node must have at least
$t$ shares to reconstruct the reading). Upon reading a value $r_i$,
sensor node $i$ chooses a random \mbox{$t - 1$} degree polynomial
$P_i(x)$ such that \mbox{$P_i(0) = r_i$}.  One may construct such a
polynomial by randomly choosing $a_{i, k}$, $\forall k \in [1, t -
1]$ and using $P_i(x) = r_i + a_{i, 1} x + a_{i, 2} x^2 + \ldots +
a_{i, t - 1} x^{t - 1}$. This is a simple and practical operation.
Each of the $p$ shares is then composed of the values $P_i(q)$ ($1 \leq q \leq
p$). Node $i$ sends then a message containing $P_i(q)$ along every
path $q$.

\vspace{1mm} \noindent {\bf Reconstruction.} In order to recover
$r_i$, one must first recover $P_i$ using polynomial interpolation and
then compute $r_i = P_i(0)$.  This operation requires at least $t$
distinct shares.  There is an infinity of $t - 1$ degree polynomials
that pass through $t - 1$ points. Thus, $t - 1$ compromised nodes
cannot guess anything about $P_i$ and $r_i$.  Also, the sink may
tolerate up to \mbox{$p - t$} non-responding nodes and still be able
to recover $r_i$.  Therefore, this scheme provides some
confidentiality and robustness against denial of service attacks even
in the presence of a few compromised nodes.

\vspace{1mm} \noindent {\bf Data aggregation.} Assume an aggregator
node along a path $q$ must fuse the readings of $i$ and $j$,
namely \mbox{$r_i = P_i(0)$} and \mbox{$r_j = P_j(0)$}. Being on
path $q$, the only data it receives is $P_i(q)$ and $P_j(q)$.  It
forwards $P_i(q) + P_j(q) = (P_i + P_j)(q)$. The same operation is
performed on the other shares of these nodes over the different
paths. By receiving $t$ samples, the sink may then recover
\mbox{$P_i + P_j$} and then \mbox{$(P_i + P_j)(0) = r_i + r_j$}. The
result also holds for multiplication and scalar division.

\vspace{1mm} \noindent {\bf Discussion.} Due to the inherent
property of secret sharing, \SMA offers very strong confidentiality.
An attacker that has not gathered at least $t$ shares cannot guess
anything about the sensor readings. The confidentiality assured by
\SMA is obtained at the cost of some overhead in data transmission
and therefore energy consumption. Upon sensing an event, $p$
messages need to be sent, each one of them being of the same size as
the original reading.\footnote{This is a well-known result of secret
sharing that can be shown easily using information theory.} This is
the main reason for which two other schemes (\DMA and \ADMA) are
proposed.

\subsection{Scheme 2: Dispersed Multipath Aggregation (\DMA)}

Information dispersal is a common technique used to introduce
redundancy and protection against Byzantine failures.  Like secret
sharing, it consists of a scheme that makes $p$ shares out of a
particular data, such that $t$ of them are needed to reconstruct the
data.  Unlike secret sharing, a data block of length $t$ is split
into $t$ pieces of length $1$.\footnote{To avoid confusion: a length
of $1$ does not mean 1 bit, but a ``unitary'' block of bits. Its
size depends on the size of an aggregate.}

\vspace{1mm} \noindent {\bf Share creation.} Each sensor is
pre-loaded with the same $t \times p$ matrix \mbox{${\bf A} = [a_{i,
q}]$}. ${\bf A}$ should be chosen in such a way that every
combination of $t$ columns should form an invertible $t \times t$
matrix. When sensing events, a sensor $i$ accumulates its readings
into an internal buffer of length $t$, considered as a vector ${\bf
R}_i = \left[\begin{array}{cccc} r_{i,1} & r_{i, 2} & \ldots & r_{i,
t}\end{array}\right]$.  This forms a block of readings.  Once the
buffer is full, the node is ready to compute $p$ different shares of
length $1$ to send along the paths. These shares are the different
elements of ${\bf M} = {\bf R}_i. {\bf A}$, where element $m_{i,q}$
of {\bf M} is given by

\begin{multline}
  \left[\begin{array}{cccc}m_{i,1}& m_{i, 2}& \ldots& m_{i, p}\end{array}\right]
  = \\
  \left[\begin{array}{cccc}r_{i,1}& r_{i, 2}& \ldots& r_{i, t}\end{array}\right]
  \cdot
  \left[\begin{array}{ccc}
      a_{1,1} & \cdots & a_{1,p} \\
      \vdots & \ddots & \vdots \\
      a_{t,1} & \cdots & a_{t,p}
    \end{array}\right].
\end{multline}

\noindent That is:
\begin{equation}
m_{i,q} = r_{i, 1}a_{1,q} + r_{i, 2}a_{2, q} + \ldots + r_{i, t}a_{t, q}.
\end{equation}

\vspace{1mm} \noindent {\bf Reconstruction.} When the sink receives
$t$ shares, it is in position of reconstructing the data. Assuming
it receives ${\bf M}_{i} = \left[\begin{array}{cccc} m_{i, q_1} &
m_{i, q_2} & \ldots & m_{i, q_t}\end{array}\right]$, readings are
obtained by resolving:

\begin{equation}
\left\{\begin{array}{c}
  r_{i, 1}a_{1, q_1} + r_{i, 2}a_{2, q_1} + \ldots + r_{i, t}a_{t, q_1} = m_{i, q_1} \\
  r_{i, 1}a_{1, q_2} + r_{i, 2}a_{2, q_2} + \ldots + r_{i, t}a_{t, q_2} = m_{i, q_2} \\
  \vdots \hspace{1cm} \vdots \hspace{1cm} \vdots \\
  r_{i, 1}a_{1, q_t} + r_{i, 2}a_{2, q_t} + \ldots + r_{i, t}a_{t, q_t} = m_{i,
  q_t}
\end{array}\right.
\label{ida-sys}
\end{equation}

This may be done using a simple Gauss elimination method or by
inverting the matrix constituted of the different $q_1, \ldots, q_t$
columns of ${\bf A}$.  If the matrix ${\bf A}$ is randomly chosen,
no known methods exist to reconstruct parts of the original data
from \mbox{$t - 1$} samples, although some correlation between the
various $r_{i, q}$ may be deduced.

\vspace{1mm} \noindent {\bf Data aggregation.} This scheme has
homomorphic properties similar to secret sharing. Given messages
$m_{i, q}$ and $m_{j, q}$ sent by nodes $i$ and $j$ on path $q$, an
aggregator node computes $m_{i, q} + m_{j, q}$.

Since one has:

\begin{equation}
m_{i, q} + m_{j, q} = \sum_{k = 1}^t (r_{i, k} + r_{j, k})a_{k, q},
\end{equation}

\noindent then, upon reception of at least $t$ such messages, the sink can
reconstitute every $r_{i, k} + r_{j, k}$ in a way similar to the
system shown in Eq.~\ref{ida-sys}:

\begin{equation}
\left\{\begin{array}{c}
  (r_{i, 1} + r_{j, 1})a_{1, q_1} + \ldots + (r_{i, t} + r_{j, t})a_{t, q_1} =
    m_{i, q_1} + m_{j, q_1} \\
  (r_{i, 1} + r_{j, 1})a_{1, q_2} + \ldots + (r_{i, t} + r_{j, t})a_{t, q_2} =
    m_{i, q_2} + m_{j, q_2} \\
  \vdots \hspace{1cm} \vdots \hspace{1cm} \vdots \\
  (r_{i, 1} + r_{j, 1})a_{1, q_t} + \ldots + (r_{i, t} + r_{j, t})a_{t, q_t} =
    m_{i, q_t} + m_{j, q_t}
\end{array}\right.
\end{equation}

\vspace{1mm} \noindent {\bf Discussion.} This scheme is space
efficient, {\it i.e.}, reconstructing $t$ readings requires only $t$
shares of the same size. Using more shares ($p>t$) allows however
for protection against \DOS attacks.

Although the scheme is more efficient in terms of overhead than \SMA, one must
keep in mind that this scheme offers a weaker confidentiality than
\SMA. Compromising nodes allows an attacker to get some information
about readings, even though partial readings cannot be
reconstructed. This provides however sufficient confidentiality for
sensor networks. One may therefore use this scheme to ensure loose
confidentiality and resistance to node failures or \DOS attacks.
Note that no heavy computations need to be performed; only the sink
has to solve the system of equations.

\subsection{Scheme 3: Authenticated Dispersed Multipath Aggregation (\ADMA)}

\SMA and \DMA as presented previously do not ensure protection from
replay attacks nor data authenticity.  A malicious attacker may
eavesdrop a sensor node and then send the messages it listened
to later.  A malicious aggregator can also send garbage bits instead of
the result of an expected computation and remain unnoticed.  Of
course, the sink may detect such an attack by performing two
reconstructions with different sets of shares and notice the results
are different. But still it cannot decide which of the results is
correct.

\vspace{1mm} \noindent {\bf Authentication.} There exist techniques
for verifiable secret sharing but they are currently impractical for
sensor networks.  For this reason, we focus on an authentication
solution for \DMA. We propose to replace the last reading of ${\bf
R}_i$ with an element that includes sequence information and depends
upon a secret shared among $i$ and the sink. Let us assume ${\bf
R}_i = \left[\begin{array}{cccc}r_{i, 1} & \ldots & r_{i, t - 1} &
h(k_i, s)\end{array}\right]$, where $h(\cdot)$ is a secure hash
function modeled as a random oracle~\cite{BelRog93}, $k_i$ is a
secret key between $i$ and the sink, and $s$ is a sequence number.

\vspace{1mm} \noindent {\bf Reconstruction.} After the
reconstruction of $\sum {\bf R}_i$, the sink just needs to verify
whether its last element is equal to $\sum h(k_i, s)$. If not, then
an aggregator node is cheating and the sink has to use another
subset of messages to reconstruct $\sum {\bf R}_i$. This is not to
be considered as a strict integrity check because the authentication
value $h(k_i, s)$ does not gather information from the readings
$r_{i, k}$.  Therefore, an attacker that has compromised $t$ nodes
might be able to reconstruct \mbox{$h(k_i, s)$} and tamper with the
data without being noticed.  But using information from $r_{i,k}$ in
the authentication value is not possible because the sink does not
know every $r_{i,k}$: it only reconstructs $\sum r_{i,k}$.  One may
notice that the use of $s$ as a simple integrity check could be
sufficient in practice. However, we opted to use a secret key $k_i$
and a hash function to make authentication values less predictable.
This not only complicates the action of tampering with the data but
it also seems to ensure a higher level of security in practice. By
using such a solution, one would lose however the space efficiency
of the scheme: reconstructing $t - 1$ readings would imply using $t$
shares instead of $t - 1$.  In order to minimize that overhead a
solution is to have blocks that contains more readings (by
increasing $t$). If this results in having more shares than the
number of paths available, one should use a large $t$ and send
multiple shares on each path.

\subsection{Summary and discussion}
\label{sec:prop-schemes:discuss}


\SMA splits each reading into a given number of shares and sends
then one share per path.  \DMA accumulates several readings in an
internal buffer before dispersing it into several shares.  It sends
one share per path.  \ADMA accumulates several readings in an
internal buffer, then inserts an authentication value into the
buffer and disperses it into several shares, possibly more than the
number of available paths.

Each of these schemes has some advantages and drawbacks. Some of
them are global to all schemes, whereas some others are specific.
First, all techniques provide resilience to unintentional failures
and \DOS attacks.  Second, all techniques hide (a varying amount of) data
from aggregator nodes, so that it is not possible for a few
compromised nodes to reconstitute the sensed data, at least
completely.

\SMA provides full confidentiality, {\it i.e.}, no information leaks
from secret shares in the sense of information theory, unless at
least $t$ of them are gathered, in which case the security
collapses completely. This strong security is obtained however at
the cost of duplicating each reading once per path.  On the other
hand, both \DMA and \ADMA are space efficient, although each share
leaks information about the readings. However, no exisitng techniques are
known  to reconstruct, even partially, sensor readings from
$t - 1$ shares.

\section{Security analysis}
\label{sec:sec-ana}

We first recall that proper lower-level mechanisms can protect a
network in the absence of node compromises~\cite{perrig04security}.
Protection from eavesdropping may be achieved with link-level
encryption.  Data tampering and packet injection are also
inefficient when facing link-level authentication and encryption.
Some physical-layer schemes and routing protocols may get around
denial of service.  But none of these techniques can protect the
network from compromised nodes.

Security analysis in our case must be done with respect to the
number of node compromises.  Each scheme disperses sensed data along
multiple paths.  Most of the time, compromising a unique node is not
sufficient for an attack to succeed: there is a threshold that
defines the minimum number of nodes an attacker need to compromise
in order to succeed in attacking.  Three fundamental questions are:

\begin{enumerate}

\item How many compromised nodes does an attacker need at best to
  eavesdrop successfully and break confidentiality for a given scheme?
  Also, which nodes should be attacked?

\item What is the minimal number of nodes an attacker need to
  compromise to inject false data into the network?  Which nodes should
  be chosen?

\item How many nodes need to be compromised at best in order
for an attacker to succeed in a \DOS attack?

\end{enumerate}

It is important to underline that an attacker might not have the
choice of which nodes to compromise. In practice, if $n$ nodes need to
be compromised for an attack to succeed, the attacker may not have
access to all of these $n$ nodes. Also, if the attacker does not
have full knowledge of the topology, it may also be difficult to
guess the interesting nodes to compromise. It may be a requirement
that an attacker needs to compromise more nodes than the theoretical
threshold.

In the following analysis, we assume without loss of generality that
only one share is sent per path.

\subsection{Eavesdropping}

In this section we analyze the resilience the proposed schemes offer to
node compromise when facing eavesdropping attacks ({\it cf.},
Section~\ref{subsec:sec-goals-threats}).

From the schemes, it appears that at least $t$ nodes are required to
be compromised in order for a node to recover data. However, there
are some subtleties.  First, such a consideration holds for \SMA
because it does not leak any information until all of the $t$ shares
are gathered. This is not the case for schemes based on information
dispersal. Second, nothing guarantees that choosing $t$ nodes from
distinct paths allows an attacker reconstructing some shares. Below
we give explanations for these two phenomena.

\subsubsection{\DMA and \ADMA information leakage}

According to information theory, due to the space efficiency of
information dispersal and since all of the $p$ shares play a
completely symmetrical role, each share contains exactly
$\frac{1}{t}$ of the readings.  Therefore, each share leaks some
information and is then a source of information for an attacker.  No
known methods are known, however, to reconstruct parts of the
readings from a subset of less than $t$ shares.

\subsubsection{Possibility of data reconstruction with $t$ compromised paths}
\label{sec:sec-ana-1}

For the reconstruction operation to work properly, share aggregates
should contain contributions from the same nodes.  However, shares
propagate on different paths, and among these paths the aggregator
nodes receive contributions from various probably different nodes.
This makes eavesdropping attacks difficult to implement.

Here is an example.  Suppose that $t = 2$ and an attacker has
succeeded in compromising two nodes $i$ and $j$ on two distinct
paths. If the shares gathered by $i$ contain contributions from,
say, nodes $k$ and $l$ and shares gathered by $j$ contain only
contributions from node $k$, then any reconstruction will be
impossible, though the attacker has compromised two distinct paths.
If shares gathered by $j$ had contained contributions from both
nodes $k$ and $l$ then the reconstruction would have been possible.

\subsection{Data tampering and packet injection}

In this section we analyze the resilience the proposed schemes offer to
node compromises when facing data tampering and packet injections
({\it cf.}, Section~\ref{subsec:sec-goals-threats}).

It is clear that an attacker that has compromised less than $t$
aggregator nodes has no effective control over the meaning of the
data it injects into the network.  Even if the attacker manages to
compromise $t$ nodes, nothing guarantees that the sink would use the
shares of all those $t$ nodes to perform a reconstruction (it may
use shares from uncompromised paths).  One can also imagine a
scenario where an attacker succeeds in compromising one or several
nodes on each possible path.  Still in this case, the attacker may
not be able to control the meaning of the injected/tampered data for
the reason described in \ref{sec:sec-ana-1}.

Finally, since \ADMA provides an authentication check, an attacker
that is not capable of reconstructing some readings (and therefore
the authentication value for each sequence) has little chance of
being able to fool the sink with tampered data.  This because (a)
the attacker does not know the expected authentication values and
(b) it will be extremely difficult for the attacker to inject a
share that would modify the readings but not the authentication
value after reconstruction.

Even without authentication, the sink may notice that some data have
been tampered if multiple reconstructions with different subsets of
shares give different outputs.  In this case, however, it cannot tell
if there is a valid subset of shares for reconstruction.

An attacker that cannot control the meaning of tampered data can at
best try to perform some kind of denial of service attack.  That is,
it may try to tamper with enough messages to make reconstruction
impossible.  In this case the security parameters will behave as
described in the following.


\subsection{Denial of service attacks}

In this section we analyze the resilience of the proposed schemes when
facing denial of service attacks ({\it cf.},
Section~\ref{subsec:sec-goals-threats}).

There are two kinds of \DOS attacks: those where attackers stop
emitting data (let us call it no-data \DOS attacks) and those where
they send garbage data (let us call it garbage-data \DOS attacks).
Garbage-data \DOS attacks are more difficult to handle.  In the
absence of data authentication, an attacker needs only to compromise
one path and send some garbage data on it.  In this case, the sink
has multiple possible outputs for reconstruction but cannot tell
which ones are valid.  In the presence of data authentication,
garbage-data \DOS attacks are indistinguishable from no-data \DOS
attacks~--- invalid reconstructions are rejected as if the wrong
share had never arrived.

No-data and garbage-data \DOS attacks in the presence of
authentication need to prevent the sink from gathering $t$ valid
shares. Therefore, an attacker needs to compromise at least $p - t +
1$ distinct paths, {\it i.e.}, in the worst case, $p - t + 1$ nodes.
If the attacker does not know the routing topology, it cannot do
anything but compromise random nodes.  Therefore, it will probably
have to compromise more than $p - t + 1$ nodes.

Let $t_e$ and $t_d$ be, respectively, the minimum number of
compromised nodes required to eavesdrop communications and the
minimum number of compromised nodes required to succeed in a \DOS
attack. From previous sections, \mbox{$t_e = t$} and \mbox{$t_d = p
- t + 1$}. Note that the higher $t_e$, the lower $t_d$. One can make
a tradeoff by choosing $\frac{t}{p}\approx\frac{p + 1}{2}$. Any
higher values would give better resistance to eavesdropping whereas
any lower values will give better resistance to \DOS attacks. Making
a relevant choice is not easy when $p$ is small ({\it e.g.}, $p =
3$).


Table~\ref{tab:bounds} summarizes the lower bounds on the number of
compromised nodes one needs to succeed under the different attacks
described above.

\begin{table}
  \caption{Lower bounds on the number of compromised nodes one needs to succeed in various
  attacks.}
  \label{tab:bounds}
  \begin{center}
    \begin{tabular}{|lcm{.5\linewidth}|}
      \hline
                        & Bound         & Comments \\
      \hline \hline
      Eavesdropping      & $t$           & Compromised shares must have the
                                           same contributing nodes. \\
      \hline
      Tampering$^\star$ & $t$            & Compromised shares must have the
                                           same contributing nodes. \\
      \hline
      \DOS attack       & $p - t + 1$   & $1$ for garbage-data \DOS
                                          attacks with \SMA or \DMA. \\
      \hline
    \end{tabular}
  \end{center}
  { \footnotesize $^\star$ Only tampering where the attacker controls
    the meaning of its falsifications is considered. }
\end{table}

\section{Further investigation}
\label{sec:further}

In this section we
first present some other approaches and compare them with our
schemes concerning both communication overhead and resistance to
attacks. We then present the implementation details and some
simulation results about the performance of our schemes.

\subsection{Comparison to other approaches}

The common insecure approach regarding data aggregation is to have
one unique tree that spans every node.  Each one of the tree's
internal nodes aggregates data from its children before forwarding
them to its parent.  With one message per node, this is the most
communication-efficient technique despite the complete lack of
security.  This aggregation method is referred to as `simple tree'
hereafter. With no overhead, one can use special encryption
techniques that provide some confidentiality and still allow for
aggregation to be performed~\cite{mykletun05efficient}.  One can
also add different authentication mechanisms, but at the cost of
larger messages~\cite{hu03secure}.  Table~\ref{tab:features}
summarizes the features of all these schemes.  As one can see,
multipath aggregation schemes provide more protection against node
compromises.

\begin{table}
  \caption{Schemes' features.}
  \label{tab:features}
  \begin{center}
    \begin{tabular}{|l|ccc|}
      \hline
                  & \multicolumn{3}{c|}{Protection from\ldots} \\
      Scheme      & eavesdropping & tampering     & \DOS attacks \\
      \hline\hline
      simple tree & no            & no            & no \\
      \hline
      Mykletun et al. \cite{mykletun05efficient}
                  & yes           & weak$^\star$  & no \\
      \hline
      Hu and Evans \cite{hu03secure}
                  & no            & yes           & no \\
      \hline
      \SMA        & yes           & weak$^\star$  & weak$^{\star\star}$ \\
      \hline
      \DMA        & yes           & weak$^\star$  & weak$^{\star\star}$ \\
      \hline
      \ADMA       & yes           & yes           & yes \\
      \hline
    \end{tabular}
  \end{center}
  { \footnotesize
    $^\star$ An attacker may alter data but not control the altered
    message's meaning. \\
    $^{\star\star}$ An attacker may succeed if she sends garbage data.
  }
\end{table}

Compared to \SMA, \DMA and \ADMA increase the delay between the time
readings are done and the time they are reported to the sink. This
is because a sensor node must temporarily fill an internal buffer
with its readings before sending them to the sink.  As an example, a
sensor node that performs \DMA and senses data every $m$ minutes
will send its messages with an interval of $t \times m$ minutes. The
first message of the sequence must wait for other readings to fill the
$t$-length buffer before the information is dispersed and sent
towards the sink.

\subsection{Communication overhead}
\label{sec:further-overhd}

A share has the same size as a unique sensor
reading; therefore, a message in such a scheme is not larger than a
message generated by the simple tree approach.

We define communication overhead as the ratio given by additional
messages sent by a node compared to the simple tree scheme.

\subsubsection{\SMA} A sensor node sends $p$ messages each time it
does a reading.  It would send one with the simple tree scheme.
Therefore, the overhead of the \SMA scheme is $p - 1$ per reading.

\subsubsection{\DMA} A sensor node with the \DMA scheme sends $p$
messages each time it does $t$ readings.  With simple tree, it would
send $t$ messages.  Thus, the overhead of \DMA is $\frac{p - t}{t}$
per reading. Note that the overhead is null when
\mbox{$p = t$}. This corresponds to the situation where all the
shares are needed to reconstruct readings. This is a consequence of
\DMA's space efficiency: when there is no data redundancy, there is
no overhead. Furthermore, this means that there is no protection
against \DOS attacks. More generally, \DMA's overhead is solely due
to data redundancy. Choosing the amount of redundancy, that is, the ratio \mbox{$\alpha
= \frac{p}{t}$} fully determines the scheme's overhead.

\subsubsection{\ADMA} A sensor node with the \ADMA scheme sends $p$
messages each time it does $t - 1$ readings.  With a simple tree
scheme, it would send $t - 1$ messages.  Therefore the overhead of
the \DMA scheme is $\frac{p - t + 1}{t - 1}$ per reading. The
minimal overhead is obtained for the minimal value of $p$, that is
$p = t$.  As with the \DMA scheme, this corresponds to the situation
where all the shares are needed to reconstruct readings.

Choosing large values of $t$ helps reducing the overhead.  One may
choose \mbox{$p = \alpha t$} for a given \mbox{$\alpha \ge 1$}. This
would ensure that at least $t (\alpha - 1)$s nodes could be
compromised and still remain robust to DoS attacks. The overhead
then becomes \mbox{$\frac{\alpha - 1 + 1/t}{1 - 1/t}$}. This means
that the larger $t$, the closer the overhead to $\alpha - 1$
(overhead for the \DMA scheme).

\vspace{1mm} \noindent {\bf Summary.} \SMA generates the highest
overhead, but \DMA and \ADMA's can be fairly reasonable depending on
the chosen parameters. A trade-off must be made between the desired
amount of data redundancy and the desired communication efficiency.
\DMA is space efficient, which means that no overhead occurs when
there is no redundancy. \ADMA is almost space efficient. On the
other hand, despite its higher overhead, \SMA may also be of
interest. It provides very strong confidentiality and may be used in
energy-unconstrained sensor networks.

\subsection{Implementation}

In this section we detail the implementation of the three proposed
aggregation schemes as well as a number of practical results. These
implementations should be seen as proof-of-concept for the
feasibility of the proposed schemes, not as complete turn-key
solutions.

\subsubsection{Setup}

Custom implementations of \SMA, \DMA, and \ADMA have been developed
for Crossbow \MICAz motes (see Section~\ref{sec:prob-form-nodes})
running TinyOS. The operations are performed over customizable prime
integer fields $GF(p)$ and we used multi-precision computation
routines from TinyECC \cite{ning05tinyecc}, which are based on
\textsc{rsaref}~\cite{rsa94rsaref}.  The source codes of the
implementations can be downloaded from
{\footnotesize \url{http://www-rp.lip6.fr/~claveiro/secure-aggreg/}}.

For the sake of simplicity and in order to isolate our results from
any bias introduced by the routing layer, we used optimized static
multipath routing. This layer uses the default TinyOS link layer,
which is not secure enough with regard to the assumptions taken in
this paper. However, thanks to TinyOS modular design, one may write
and use his own layers for routing and secure-link establishment
without being intrusive.

\begin{figure}[t]
  \begin{center}
    \includegraphics[width=\linewidth]{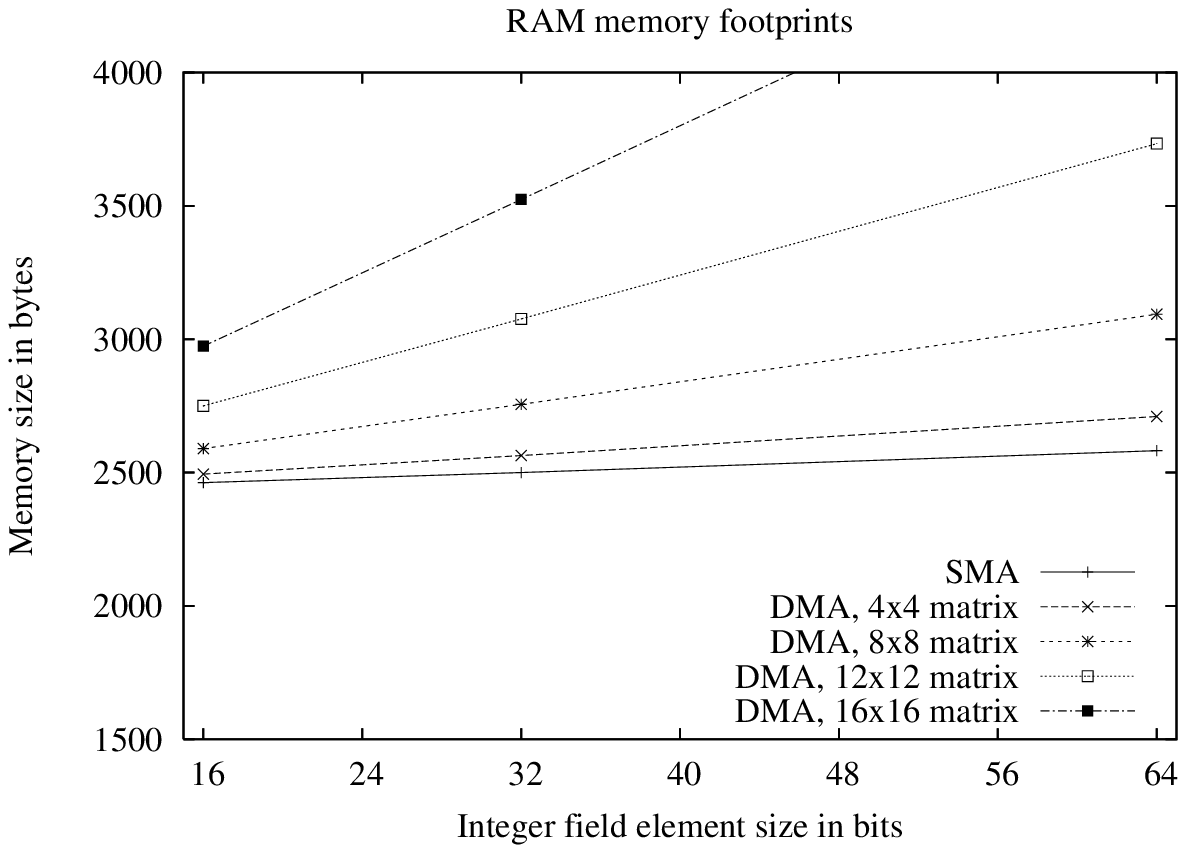}
    \includegraphics[width=\linewidth]{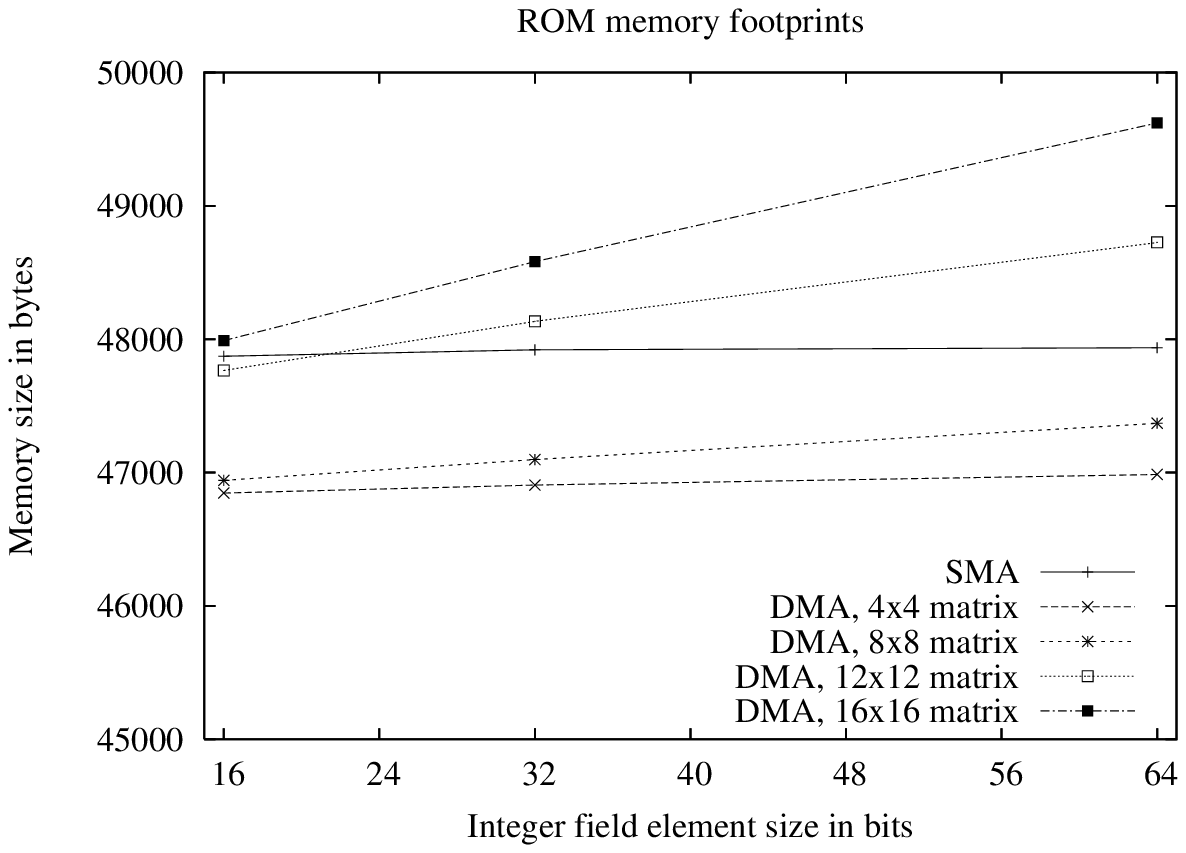}
  \end{center}
  \caption{{\it Implementations' memory footprints.}  For the sake of
    readability, \mbox{\ADMA's} footprints are not depicted here.  Their RAM
    consumptions are identical to those of \DMA.  ROM occupancy is
    slightly bigger due to the required extra code for authentication
    values computation.}
  \label{fig:code-size}
\end{figure}

Once compiled, many parameters impact memory occupancy.  Let us
consider the size of a $GF(p)$ integer or the information dispersal
${\bf A}$ matrix size.  Figure~\ref{fig:code-size} presents the
memory footprints for some typical parameters.  \SMA's footprint is
rather good whatever integer field is used (about the half of a
\MICAz's RAM, for instance).  \DMA and \ADMA's footprints are very
sensitive to the size of the information dispersal ${\bf A}$ matrix.
This matrix determines the maximum number of shares $p$ and
threshold $t$ of \DMA and \ADMA schemes.  For given $p$ and $t$
parameters one needs a \mbox{$t \times p$} matrix.  Some big values,
such as \mbox{$16 \times 16$} matrices with 64 bits integers do
not fit into a \MICAz mote.  Other values are however fairly
reasonable with respect to memory occupancy.

The time required for the nodes to perform operations such as share
creation and aggregation is negligible and has never been an issue
during tests.

The aggregation processes work as follows.  Nodes sense some data at
regular intervals and push shares toward the sink using a sequence
number.  An aggregator node only aggregates shares having the same
sequence number.  When an aggregator node receives a share, it
stores the share in a buffer and waits for other shares with the
same sequence number.  If other shares arrive, the node aggregates
them and keeps waiting until a new share with a higher sequence
number arrives or a timer goes off.

\subsubsection{Experimentations}

We performed both real experiments and simulations using the
implementation described above. Experiments were done at small scale
(six nodes and three-path topologies) to test the practicality of
the schemes. In order to stress the implementation, we performed
simulations using \TOSSIM. \TOSSIM is a sensor network simulator
that compiles directly from TinyOS code and simulates the TinyOS
network stack at the bit level. This has the advantage of perfectly
modeling the behavior of the implementation.

\begin{figure}[t]
  \centering
  \includegraphics[width=\linewidth]{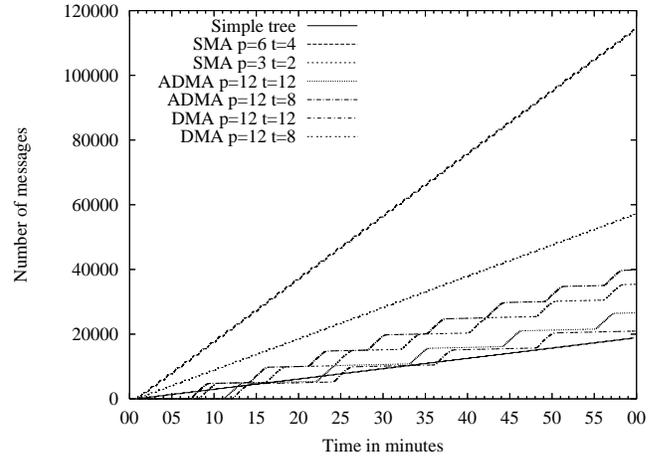}
  \caption{Number of sent messages \wrt schemes.}
  \label{fig:messages}
\end{figure}

We measured the number of messages for different parameters and
schemes of the implementations.  We used five random topologies of
forty nodes with one hour of simulation time.  These topologies use
sink-rooted node-disjoint trees to perform multipath routing.  Two
topologies have four paths, the others have respectively three, six,
and eight paths. Note that the number of paths does not impact on
the measured number of sent messages, which is solely influenced by
parameters $p$ and $t$. When there are more shares than available
paths, some paths carry multiple shares. When there are more paths
than available shares, some paths are unused.
Figure~\ref{fig:messages} represents the total number of sent
messages among the five topologies with respect to simulation time.
We compare this number with the number of messages for the simple
tree scheme with the previously described aggregation technique.

We can observe some predictable properties of the schemes.  As
previously analyzed, \SMA's overhead is the highest one, depending
solely on the $p$ parameter.  Therefore, \SMA roughly needs $p$
additional messages compared to the simple tree scheme.  We can also
see that the overhead of information dispersal based schemes depends
on $\alpha = \frac{p}{t}$. It is not a surprise that \ADMA and \DMA
exhibit similar performance, with \DMA having a slightly better
overhead. Overheads are however a bit higher than computed in
section \ref{sec:further-overhd}.  As an example, \ADMA with
\mbox{$p = 12$} and \mbox{$t = 8$} has an overhead of $1$ instead of
the predicted $0.7$.  Also, \ADMA with \mbox{$p = t = 12$} exhibit a
small overhead of $0.25$ instead of about $0.1$. This is due to the
practical considerations that make implementations miss some
aggregation opportunities.

\subsection{Cost vs. Security}

Security necessarily implies a cost with regard to some metric. Some
schemes generates overhead in terms of communications, others in
terms of CPU consumption, etc. What is important to define is a
solution that leads to the required level of security at the cost at
an acceptable overhead. In this way, our proposals are very
promising. Indeed, by using the proposed schemes, a network
tolerates multiple compromises without jeopardizing confidentiality,
authenticity, and availability. Thus, the overheads generated by
\SMA, \DMA, and \ADMA are acceptable. Furthermore, one can customize
the overhead by adjusting the different parameters of the schemes.
Depending on the amount of resources allocated to security, one may
trade-off some security for some communication efficiency.

\section{Conclusion and future works}
\label{sec:concl}

In this paper we proposed three schemes to secure data aggregation using
multipath routing.  They are based on secret sharing and information
dispersal.  In the proposed schemes, sensors split their readings
into several shares and distribute them among several disjoint
paths. Upon reception of a minimum number of shares, the sink can
reconstruct the aggregated value.

Depending on the scheme and its parameters, these techniques provide
varying levels of resistance to \DOS attacks, eavesdropping, and data
tampering. By using secret multipath aggregation, one can guarantee
that a subset of compromised paths cannot reveal/leak any information
about the readings. This is at the cost of some overhead. By using
dispersed multipath aggregation, one has an optimal overhead but
achieves lower levels of confidentiality. Depending on the
application or scenario, one approach offers more advantages over
the other.

To the best of our knowledge, the three proposed schemes
are the first to address node compromises for aggregation schemes in
sensor networks using multiple paths. Future work concerning these
schemes includes modeling the security parameters' statistical
behavior under the contingency of random node compromises.  It is also
possible to   generalize  and apply these schemes to contexts  other
than sensor networks.


\IEEEtriggeratref{5}

\bibliographystyle{IEEEtran}
\bibliography{IEEEabrv,wsn.v04}

\end{document}